\documentclass[conference]{IEEEtran} 
\usepackage{amsmath}
\usepackage{amsmath,amssymb,amsfonts}
\usepackage{cite}
\usepackage{mathtools}
\usepackage[binary-units=true]{siunitx}
\DeclareSIUnit{\belmilliwatt}{Bm}
\usepackage{amsfonts}
\usepackage{amssymb}
\usepackage{graphicx}
\usepackage{dsfont}
\usepackage{subcaption}
\usepackage{comment}
\usepackage{algorithm}
\usepackage{algpseudocode}
\usepackage{mathtools}

\usepackage{textcomp}
\usepackage[export]{adjustbox}
\usepackage{float}
\usepackage{booktabs}
\usepackage{multicol}
\usepackage{xparse}
\usepackage[left=1.62cm,right=1.62cm,top=1.9cm]{geometry}
\usepackage{color,soul}
\usepackage[bottom]{footmisc}
\usepackage[binary-units=true]{siunitx}
\DeclareSIUnit{\belmilliwatt}{Bm}
\DeclareSIUnit{\dBm}{\deci\belmilliwatt}
\DeclareSIUnit[per-mode=symbol,per-symbol=p]{\Bps}{\byte\per\second}

\sethlcolor{yellow}
\usepackage{algpseudocode}

\usepackage{hyperref}

\makeatletter
\def\BState{\State\hskip-\ALG@thistlm}
\makeatother

\IEEEoverridecommandlockouts
\begin{document}

    \title{Distributed Learning for Reliable and Timely Communication in 6G Industrial Subnetworks}

\author{
\IEEEauthorblockN{Samira Abdelrahman\IEEEauthorrefmark{1}, Hossam Farag\IEEEauthorrefmark{2}\IEEEauthorrefmark{1}, and  Gilberto Berardinelli\IEEEauthorrefmark{2}}
\IEEEauthorblockA{
\IEEEauthorrefmark{1} Department of Electrical Engineering, Aswan University, Egypt\\
\IEEEauthorrefmark{2}Department of Electronic Systems, Aalborg University, Denmark \\
Email: sma@asw.edu.eg,  \{hmf, gb\}@es.aau.dk}
}

	\maketitle
	
	\begin{abstract}  
Emerging 6G industrial networks envision autonomous in-X subnetworks to support efficient and cost-effective short range, localized connectivity for autonomous control operations. Supporting timely transmission of event-driven, critical control traffic is challenging in such networks is challenging due to limited radio resources, dynamic device activity, and high mobility. In this paper, we propose a distributed, learning-based random access protocol that establishes implicit inter-subnetwork coordination to minimize the collision probability and improves timely delivery. Each subnetwork independently learns and selects access configurations based on a contention signature signal broadcast by a central access point, enabling adaptive, collision-aware access under dynamic traffic and mobility conditions. The proposed approach features lightweight neural models and online training, making it suitable for deployment in constrained industrial subnetworks. Simulation results show that our method significantly improves the probability of timely packet delivery compared to baseline methods, particularly in dense and high-load scenarios. For instance, our proposed method achieves $21\%$ gain in the probability of timely packet delivery compared to a classical Multi-Armed Bandit (MAB) for an industrial setting of $60$ subnetworks and 5 radio channels

	\end{abstract}
\section{Introduction}\label{sec:intro}
The modular nature of next-generation industrial systems demands the replacement of inflexible wired infrastructures with reliable wireless communication that enables flexibility, modularity, and  self-reconfigurability at the field level. This represents a key objective of 6G, particularly within the context of in-X subnetworks situated at the edge of the 6G “network of networks” architecture~\cite{IN-X1}. In-X subnetworks are localized, short range radio cells installed in robots and production modules, operating in a semi-independent fashion while benefiting from the parent 6G network when available~\cite{IN-X1}. Within this framework, Local Access Points (LAPs) manage wireless connectivity within a subnetwork and act as intermediaries that aggregate information from sensors and relay critical updates to a Central Access Point (CAP) in response to abnormal or safety-critical events. Despite their promise, in-X subnetworks face significant challenges pertaining to support low latency and reliable communication in time-critical industrial applications. Specifically, event-triggered, deadline-sensitive traffic generates abrupt spikes in channel access demand, especially when multiple devices react simultaneously to a shared abnormality. For in-X subnetworks, the high device density and node mobility exacerbate the dynamics of traffic, channel fading, and spatial interferences. These peculiar characteristics strain conventional scheduling and channel access mechanisms and demand robust, adaptive strategies. 

In the context of industrial wireless networks, industrial protocols such as WirelessHART, ISA100.11a, and WIA-PA predominantly rely on Time-Division Multiple Access (TDMA) to manage medium access~\cite{deadline}. While TDMA offers predictable latency and avoids collisions by design for periodic traffic, it cannot accommodate the bursty and asynchronous nature of emergency traffic. As such, event-triggered packets may be significantly delayed while waiting for a scheduled slot, leading to deadline violations and missed control opportunities~\cite{mixed}. On the other side,  in Random Access (RA) schemes such as ALOHA, devices are allowed to transmit sporadic traffic without explicit coordination. While RA supports spontaneous transmissions, the uncoordinated access approach leads to high collision rates, especially in dynamic and dense environments where many devices may attempt simultaneous access in response to a certain event. Moreover, both access schemes are not originally designed to support the unique characteristics of in-X subnetworks, including high mobility, dense device deployments, and strict timing constraints. In that context, intelligent access techniques, adopting Machine Learning (ML), have been shown as a promising solution where devices can, in an autonomous and distributed manner, learn optimal channel access strategy in response to environmental context and past experience.

In recent years, several works have been introduced targeting interference management and radio resource allocation in 6G in-X subnetworks~\cite{sub3, sub4, sub5, sub6, sub7}.  For instance, the authors in~\cite{sub3, sub4} use graph neural networks to model the spatial dependencies in wireless subnetworks and optimize power control policies. While these techniques effectively exploit structural information, they primarily target average performance metrics such as throughput and energy efficiency, and are not inherently designed to handle deadline-sensitive or alarm-driven traffic, which requires real-time responsiveness. Other recent studies~\cite{sub5, sub6, sub7} have employed multi-agent reinforcement learning  to address dynamic spectrum access and resource allocation in mobile and dense in-X subnetworks. These works demonstrate that distributed learning agents can adapt to local interference patterns and jointly optimize long-term rewards. However, they typically rely on reward structures and training objectives focused on aggregate utility, fairness, or throughput, without explicitly encoding temporal deadlines or reliability constraints. As a result, they may converge to policies that improve overall network efficiency while failing to prioritize urgent, time-critical transmissions.  Different works have proposed improved communication protocols to enhance delay and reliability in industrial networks. In~\cite{RA2}, the authors proposed  a multi-agent deep reinforcement learning (DRL) framework for random access (RA) in massive IoT environments, adopting a centralized training and decentralized execution paradigm. However, the proposed random access policy fails to converge in scenarios where device identifiers are unavailable, limiting its applicability in decentralized and anonymous industrial deployments. A Multi-Armed Bandit (MAB)-based framework is introduced in~\cite{NP} that enables indirect coordination among devices during alarm transmission. However, the proposed method works under the assumption of no more than three devices are activated at the same time, which is impractical in realistic industrial settings and impose strict limitation on its applicability in dense industrial in-X subnetworks. 

In this paper, we propose an efficient distributed learning-based RA protocol to support time-critical communication in 6G industrial in-X subnetworks. We focus on a scenario where a number of LAPs are triggered simultaneously to transmit critical alarms to the CAP within a strict deadline constraint. Each LAP learns a context-aware access strategy by selecting an access configuration (i.e., transmission over selected channels) based on the knowledge of the real-time contention state  which is conveyed by a broadcast signal, namely contention signature (CS) signal, transmitted by the CAP. The RA strategy is formulated as an optimization problem with the objective to maximize the probability that an alarm is delivered successfully to the CAP within its predefined deadline. The formulated problem is solved at each LAP by a deep neural network (DNN) that processes the received CS signal to guide the LAPs towards optimal access decisions, enabling implicit inter-LAP access coordination.  We adopt an online training approach where each LAP updates the DNN parameters and adjust their access decisions based on the received context signal. Simulations results show that our proposed approach improves the the probability of timely delivery by $21\%$ with respect to a baseline MAB method.

The rest of the paper is organized as follows.
Section~\ref{system-model} presents the system model. The proposed distributed learning framework is introduced in Section~\ref{proposed}. Performance evaluations are presented in Section~\ref{results}, and finally the paper is concluded in Section~\ref{sec:conclusions}
    \begin{figure}[t!] 
		\centering
		\includegraphics[width= 0.9\linewidth]{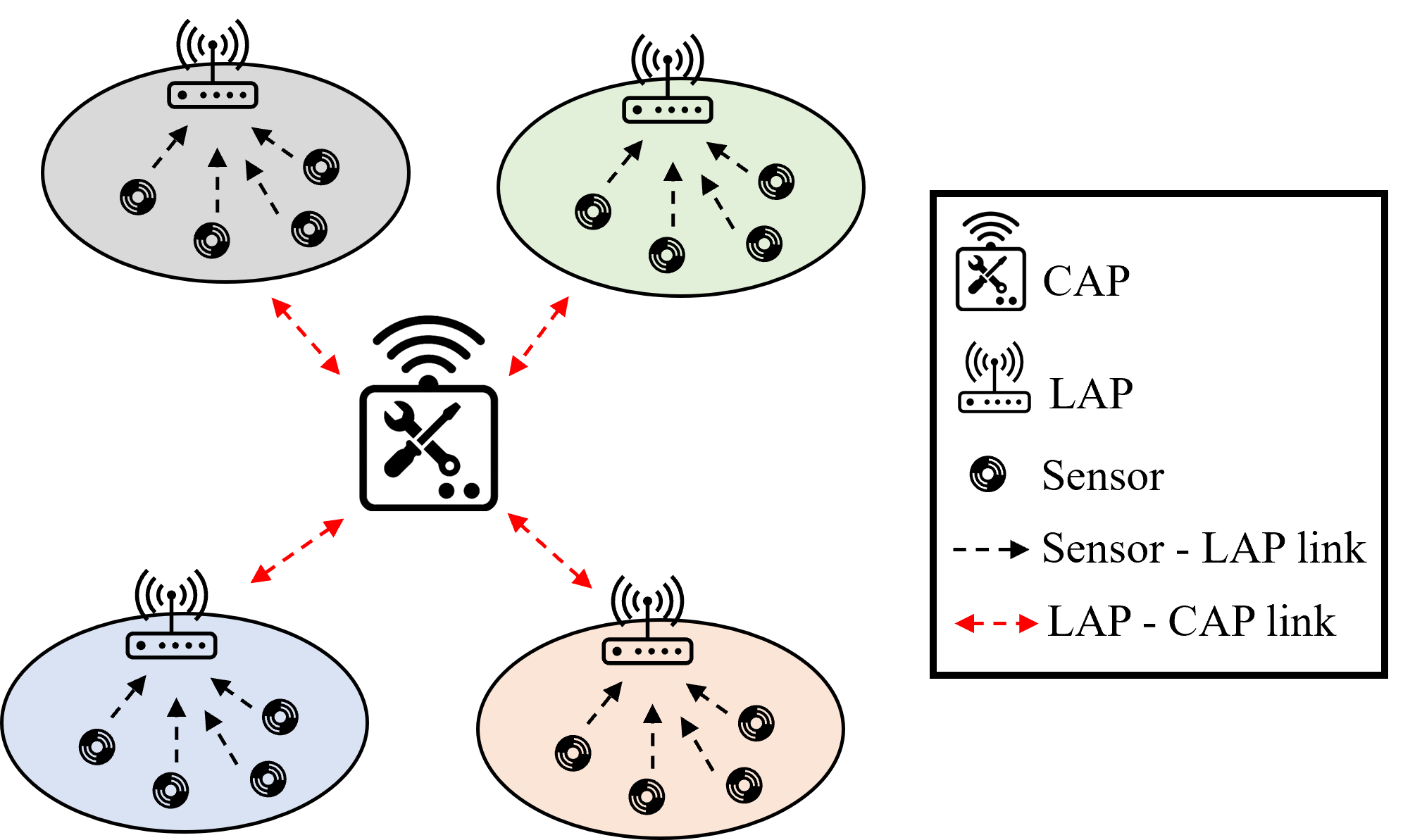}
		\caption{Network model of an industrial scenario comprising a set of mobile subnetworks and one CAP.  \label{system}}
		\vspace{-7mm}
	\end{figure}

\section{System Model and Problem Formulation}
\label{system-model}
We consider a smart manufacturing facility comprising a set $\mathcal{K}$ of mobile subnetworks as shown by Fig.~\ref{system}. Each subnetwork includes a set of sensors and one Local Access Point (LAP) which provides radio, as well as simple processing capabilities. In addition, there is one AP acts as a central controller (CAP) that is capable of monitoring the industrial process and issuing control commands to different actuators to ensure the stability and safety of the industrial process. Each subnetwork moves at a speed $v$ where sensors periodically collect measurements (e.g., temperature, camera data, ..., etc.) and communicate them to their corresponding LAPs via TDMA, avoiding intra-subnetwork interference. The LAPs communicate with the CAP only in case of abnormal or emergency situations, which are detected based on the received sensor measurements. The setup represents typical industrial use-case where a set of autonomous mobile robotic units equipped with in-X cells are deployed to perform real-time inspection, safety checks, and environment monitoring across different segments of hazardous or hard-to-reach areas that are unsafe for human operators~\cite{6Gshine}. The LAPs share a bandwidth of $M$ orthogonal channels, with $M<<K$, to concurrently transmit the generated updates to the CAP. At each time slot, a random set $\mathcal{\grave{K}}\subseteq \mathcal{K}$ of LAPs, with cardinality $|\mathcal{K}|$, are activated with probability $\grave{p}$ to transmit information (critical) updates to the CAP which in turn utilizes the received updates to stabilize the industrial process. Each activated LAP transmits the queued updates in its output buffer following First-In-First-Out (FIFO) discipline. Successful transmissions are acknowledged by the CAP via an error-free channel, and a failed update is retransmitted until is successfully delivered. A received update is regarded as valid if it is delivered within a predefined deadline $D$ since its generation, otherwise it becomes useless to the CAP. The goal is to ensure that active LAPs transmit their updates successfully within the deadline $D$ on at least one channel. 
Once activated, each LAP $n\in \mathcal{\grave{K}}$ selects an access configuration $\boldsymbol{b}_n=[b_{n,1}, b_{n,2}, ..., b_{n,M}]^T$ where $b_{n,m}=\{0,1\}$ is the channel selection indicator where we have
\begin{equation}
    b_{n, m} =
\begin{cases}
1, & \text{if LAP } n \text{ decides to transmit over channel } m, \\
0, & \text{otherwise}.
\end{cases}
\end{equation}
Let the matrix $\boldsymbol{B}\in\{0,1\}$ represents all the access configurations  of active LAPs, which has a size of $M\times |\mathcal{\Grave{K}}|$. Each column in $\boldsymbol{B}$ represents the access configuration of each activated LAP. A successful reception of the  update by the CAP is denoted by the indicator $\delta(\mathcal{\grave{K}}, \boldsymbol{B})$ which is given as
\begin{equation}\label{sucess-ind}
\delta(\mathcal{\grave{K}}, \boldsymbol{B})=
    \begin{cases}
   1, & \exists m \in \{1, ..., M\}: \sum_{n \in \mathcal{\Grave{K}}} b_{n,m} = 1, \\    
   0, & \textrm{otherwise.}
    \end{cases}
\end{equation}
Not that for $\delta(\mathcal{\grave{K}}, \boldsymbol{B})$ in \eqref{sucess-ind}, we consider a simple collision channel, i.e., a transmission fails only due to collision, which a commonly adopted assumption~\cite{NP}. With a total bandwidth of $M$ channels, each LAP $n$ can select from $2^M$ access configurations. All possible access configurations can be defined by the the matrix $\grave{\boldsymbol{B}}=[\grave{\boldsymbol{b_1}}, \grave{\boldsymbol{b_2}}, ..., \grave{\boldsymbol{b_{2^M}}}]\in \{0,1\}$, of size $M\times 2^M$, where $\grave{\boldsymbol{b_i}}\neq \grave{\boldsymbol{b_j}}, \forall i\neq j$. We denote the probability of an active LAP $n$ to choose an access configuration $\grave{\boldsymbol{b_i}}$ as  $\psi_{n}(\grave{\boldsymbol{b_i}})$ where we have $\sum_{i=1}^{2^M} \psi_{n}(\grave{\boldsymbol{b_i}}) = 1,  \forall n \in \mathcal{\Grave{K}}$. In addition, we define the matrix $\boldsymbol{\Psi}$, of size $|\mathcal{K}|\times 2^M$, whose $n^{\textrm{th}}$ row represents the elements of the set $\{\psi_{n}(\grave{\boldsymbol{b_i}})|\forall i=1, ..., 2^M\}$. Based on the law of total probability, we can represent the successful transmission probability as 
\begin{equation}\label{success_prob}
\Lambda(\boldsymbol{\Psi}) =
\sum_{\mathcal{\grave{K}} \in \mathcal{P}(\mathcal{K})}
\grave{p}
\sum_{A \in \{0,1\}^{M \times |\mathcal{\grave{K}}|}}
\left(\delta(\mathcal{\grave{K}}, \boldsymbol{B}) \prod_{n \in \mathcal{\grave{K}}} \psi_{n}(\boldsymbol{b_n}) \right).
\end{equation}

\begin{figure}[t!] 
		\centering
		\includegraphics[width= 0.8\linewidth]{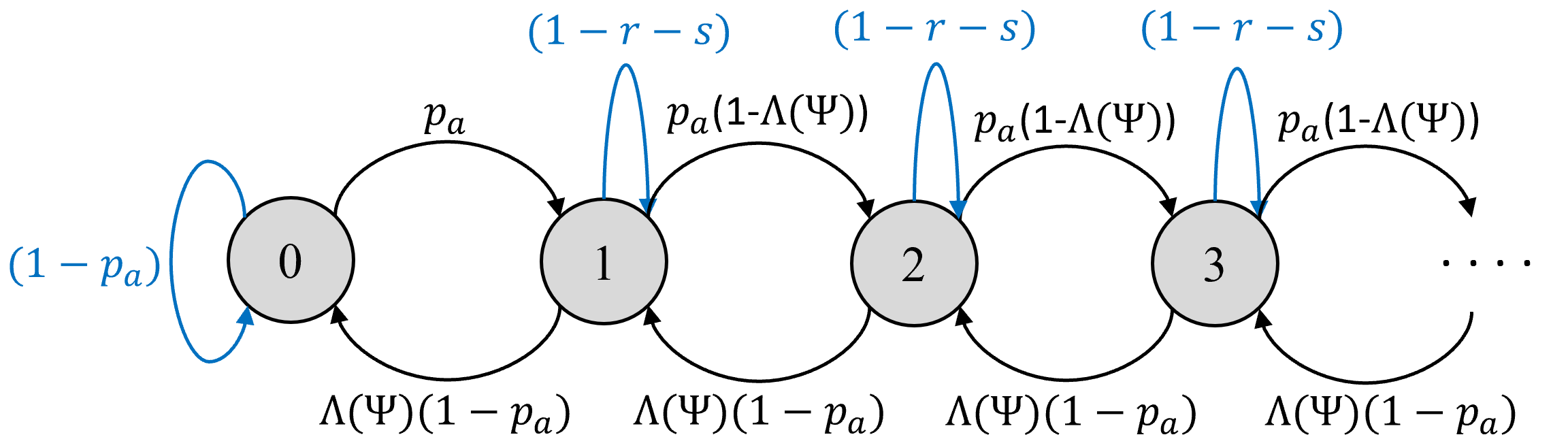}
		\caption{DTMC model of the queue size of an arbitrary LAP.  \label{DTMC-deadline}}
		\vspace{-5mm}
	\end{figure}

Next, we drive the deadline violation probability $P_D$, which is the probability that the transmission delay of an alarm update exceeds a predefined deadline $D$. Denoting $T$ as the random variable representing the delay of an update, then we have  $P_D =\mathrm{Pr}\{T>D\}$. In order to calculate $P_D$, we derive the distribution of the steady-state delay $T$ of an update.
Each node attempts to transmit the Head of the Line (HoL) update until it is successfully delivered with probability $\Lambda(\Psi)$~\cite{geo}. Hence, the service time $A_j$ of the $j$th update is random variable following a geometrically distribution with parameter $\Lambda(\boldsymbol{\Psi})$. Consider a tagged update that arrives while there are $L$ updates already in the buffer.
The delay of this update is the random sum $T = A_1 + A_2 + ... + A_L$, where $A_j,\, j = 1, 2, ...., L,$ are independent and identically distributed geometric random variables with parameter $\Lambda(\boldsymbol{\Psi})$. The Discrete-Time Markov Chain (DTMC) in Fig.~\ref{DTMC-deadline} shows the queue evolution of an arbitrary LAP $n$, where $r= p_a(1-\Lambda(\Psi))$ and $s =\Lambda(\boldsymbol{\Psi})(1-p_a)$, where the update arrival process is modeled as independent Bernoulli process with probability $p_a$.
Let $Q_j$ be the steady-state probability of having $j$ updates in the queue of $n$, then using the balance equations of the DTMC in Fig. \ref{DTMC-deadline}~\cite{geo}, we obtain the following 

\begin{equation}
Q_j = 
    \begin{cases}
\rho^{j-1} Q_1 & j\geq 1\\
\frac{\Lambda(\boldsymbol{\Psi}) (1-p_a)}{p_a} Q_1 & j=0
    \end{cases}
\end{equation}
where $\rho = \frac{r}{s}$, $Q_1 = \frac{p_a(1-\rho)}{\Lambda(\Psi)}$. Applying the convolution property~\cite{conv}, the probability generating function $G_T(z)$ is
\begin{equation}\label{PGF}
\begin{split}
 G_T(z) &= \sum_{j=1}^\infty \left(\frac{\Lambda(\Psi) z}{1-(1-\Lambda(\boldsymbol{\Psi}))z}\right)^j (1-\rho)\rho^{(j-1)}  \\
 & \approx \frac{q_s (1-\rho)z}{1-(1-\Lambda(\boldsymbol{\Psi})(1-\rho))z}.
\end{split}
\end{equation}
Based on \eqref{PGF}, the PDF of $T$ is obtained as
\begin{equation}\label{PDF_T}
f_T(t) = \Lambda(\boldsymbol{\Psi}) (1-\rho)(1-(\Lambda(\boldsymbol{\Psi})(1-\rho)))^{(t-1)}
\end{equation}
and the deadline violation probability $P_D$ is obtained as
\begin{equation}
\begin{split}\label{violation}
P_D &= \mathrm{Pr}\{T>D\} = 1-\mathrm{Pr}\{T\leq D\} \\
 &= 1-\sum_{i=1}^D f_T(i) = (1-(\Lambda(\boldsymbol{\Psi})(1-\rho)))^D.
\end{split}
\end{equation}
We define our optimization problem whose solution gives $\boldsymbol{\Psi}^*$ one of the access configurations that minimizes the deadline violation probability $P_D$. Based on \eqref{violation}, such optimization problem  can be formulated as
\begin{subequations}\label{problem}
\begin{align}
\min_{\boldsymbol{\Psi}} \quad & P_D(\Lambda(\Psi)), \label{eq:4a} \\
\text{subject to} \quad & \sum_{i=1}^{2^M} \psi_{n}(\grave{\boldsymbol{b_i}}) = 1, \quad \forall n\in \mathcal{\grave{K}}, \label{eq:4b} \\
& \boldsymbol{\Psi} \in [0,1]^{|\mathcal{K}| \times 2^M}. \label{eq:4c}
\end{align}
\end{subequations}
To solve the formulated problem in \eqref{problem}, a full knowledge about which LAPs are active at a given time ($p_{\mathcal{\grave{K}}}(\mathcal{\grave{K}})$), which is infeasible in practice. In addition, as we have $|\mathcal{\grave{K}}|=1$ or $|\mathcal{K}|>M$, the problem becomes NP-hard~\cite{NP}. In the following, we introduce an online learning approach that can solve the problem without explicit knowledge about  $p_{\mathcal{\grave{K}}}(\mathcal{\grave{K}})$. 
\section{The Proposed DNN-Based Channel Access Protocol}
\label{proposed}
In this work, we introduce our proposed transmission protocol, which is based on a distributed learning approach. In this approach, LAPs tend to learn the optimal access configuration to communicate with the CAP with the ultimate goal of maximizing $\Lambda(\boldsymbol{\Psi})$, i.e., minimizing $P_D$. Each LAP independently trains a deep neural network (DNN) using online learning to optimize its channel selection strategy without requiring explicit coordination.
\subsection{The Contention Signature Signal}
Each active LAP $n\in \mathcal{\grave{K}}$ first transmits a pilot signal to the CAP, indicating that abnormal activity has been detected and a stream of updates need to be communicated. The pilot signal comprises $M$ symbols each is transmitted on a separate channel. The CAP receives the pilot signals from all active LAPs and aggregates them into one broadcast signal, referred as the contention signature (CS) signal and sends it back to LAPs in the next time slot. 

The CS signal implicitly informs LAP agents that there are a set of contending devices (i.e., indicates the current contention level) which can be utilized, in a distributed manner, to learn an optimal access configuration to minimize $P_D$. The aggregated pilot signal $y$ received by the CAP can be expressed as
\begin{equation}\label{pilot}
\boldsymbol{y} = \sum_{n \in \mathcal{\grave{K}}} \sqrt{\eta} \operatorname{diag}(\boldsymbol{h}_n) \boldsymbol{x}_n + \sigma,
\end{equation}
where $\eta$ is the average signal-to-noise ratio (SNR), $\boldsymbol{h}_n \in \mathbb{C}^{M\times 1}$ represents the $M$ channel coefficients between LAP $n$ and the CAP, $\boldsymbol{x}_n = [x_{1,n}, x_{2,n}, ..., x_{M,n}]^T\in \mathbb{C}^{M\times 1}$ is the pilot sequence from LAP $n$ and $\sigma$ is the Additive White Gaussian Noise (AWGN) with normalized unit power. The CAP then transmits back the CS signal to all LAPs. The received CS signal $\boldsymbol{y}_n$ is given by
\begin{equation}\label{broadcast}
\boldsymbol{y}_n =  \sqrt{\eta} \operatorname{diag}(\boldsymbol{h}_n) \boldsymbol{y} + \grave{\sigma},
\end{equation}
where $ \grave{\sigma}$ is the normalized AWGN. For the sake of simplicity, we assume the same $\eta$ for the CAP. The received CS encapsulates useful information about the network state, including: the presence of other active LAPs, the channel conditions experienced by those LAPs and implicit indicators of device density and interference.
\subsection{Reward and Action-Value Function}
 Each agent (LAP) receiving the broadcast signal $\boldsymbol{y}_n$ selects an action, which refers to choosing an access configuration from the available $2^M$ in $\boldsymbol{\grave{B}}$. As mentioned earlier, a successfully received update is  regarded as valid only if its received within the predefined deadline $D$ in at least one of the available $M$ channels. An agent will get a reward $r(\boldsymbol{b}_n)=1$ if the update is received within $D$, and $r(\boldsymbol{b}_n)=-1$ otherwise. The active agents attempt to maximize their long term reward by learning to maximize the action-value function $\boldsymbol{V}(\boldsymbol{y_n}, \boldsymbol{\grave{b_i}}), \forall i=1, 2, ..., 2^M$, which represents the action $\boldsymbol{\grave{b_i}}$ given the network state $\boldsymbol{y_n}$. We utilize the $\epsilon$-greedy algorithm by selecting the action $\boldsymbol{b}_{n} = \arg\max_{\boldsymbol{\grave{b}}_n \in \{0,1\}^M} \boldsymbol{V}(\boldsymbol{y_n}, \boldsymbol{\grave{b_i}})$ with probability $1-\epsilon$ while selecting a random action with probability $\epsilon$. Due to the dynamic nature of industrial subnetworks, including the channel fading, mobility, and noise, the LAP agents would receive different CS signal $\boldsymbol{y}_n$ each time they become active. This would lead to having infinite number of states, which makes direct action-state mapping (i.e., separate action for each broadcast signal) is infeasible. Therefore, we consider a parameterized action-value function where the LAP agents tune its parameters to match $\boldsymbol{V}(\boldsymbol{y_n}, \boldsymbol{\grave{b_i}})$ with the reward obtained after observing the
feedback from the CAP. The parameterized action-value function is expressed as $\hat{\boldsymbol{V}}(\boldsymbol{y_n}, \boldsymbol{\grave{b_i}}, \boldsymbol{w})$, where $\boldsymbol{w})$ is the tunable-weights vector of the DNN layers.
\subsection{Architecture and Training of the DNN}
 The input to the DNN is the received CS signal. This signal is then passed to two fully connected hidden layers, each with size $q$, followed by a dense output layer. The two hidden layers are sufficient to approximate any smooth function to an arbitrary degree of accuracy, and are capable of generalizing the DNN to different network conditions and traffic scenarios~\cite{DNN-hidden}. The DNN produces $2^M$ parameterized action values ($\hat{\boldsymbol{V}}$) at its output. To enhance training efficiency and stability, the DNN incorporates: (a) the ReLU activation function, which serves as a computationally efficient alternative to sigmoid and tanh functions and helps mitigate vanishing gradient issues; and (b) the  Root Mean Square Propagation (RMSProp) optimizer, which adapts the learning rate of each weight individually by maintaining a moving average of the squared gradients. Each agent will collect and store a tuple of $\{\boldsymbol{y_n}, \boldsymbol{b}_n,  r(\boldsymbol{b}_n)\}$ in its local memory, of size $S$. Once the memory is full, the LAP agent will discard the oldest tuple to store a new one. In that way, the agents maintain the recent information to train the DNN and perform fine-tuning of the weights without the need to store a full history in their limited memory. 
 
 The LAP agents follow an online training approach, by performing a random sampling of a mini-batch of size $B$ of the stored data and provide it as an input to the DNN to update the weight vector $\boldsymbol{w}$ by minimizing a cost function $J(\boldsymbol{w})$
 \begin{equation}
J_n(\boldsymbol{w}) = \frac{1}{B} \sum_{j=1}^{B} \left[ r_j(\boldsymbol{b}_n) - \boldsymbol{\hat{V_j}}(\boldsymbol{y_n}, \boldsymbol{\grave{b_i}}, \boldsymbol{w}) \right]^2,
\end{equation}
where $r_j$ and $\hat{\boldsymbol{V_j}}$ denote the reward and the action-value function of the $j$-th sample, respectively.  The random selection of mini-batches improves the model generalization and help to mitigate overfitting to specific short-term trends. Furthermore, training the DNN with mini-batches helps in reducing the variance in the DNN updates, leading to a more stable and smooth learning. It is worth noting that the selection of an appropriate mini-batch size $B$ is critical; a small $B$ can introduce high variance and cause oscillations in the learned protocol, while a large $B$ increases computational overhead and the risk of overfitting to the sampled mini-batch.
\subsection{Design of the Transmission Protocol}
Each LAP maintains one of two transmission states; 1) normal state, 2) alarm state. In the normal state, the LAPs receive and process updates from their connected sensors while having no transmissions to the CAP. In the alarm state, the active LAPs stop receiving data from the sensors and transmit alarms to the CAP according to the following procedures. Once an alarm is detected, the active LAP transmits a pilot signal to the CAP consisting of $M$ symbols corresponding the $M$ available channels. The CAP receives the pilot signals from active LAPs and transmits back the CS signal. The active LAPs receive the CS signal, feed it to their DNN and select their access configurations. Each active LAP then transmits the alarm to the CAP according to the selected access configuration and sets two timers; one for the deadline $D$ (referred as $T_D$) and one for the ACK response (referred as $T_{ACK}$). If $T_{ACK}$ expires, the LAP retransmits the alarm until it receives an ACK from the CAP. If an ACK is received from the CAP before the $T_D$ expires, the transmission is deemed successful, and the LAP sets $r(\boldsymbol{b}_n)=1$. However, if the $T_D$ expires and no ACK is received, the LAP sets $r(\boldsymbol{b}_n)=-1$. 
\section{Performance Evaluation}
\label{results}
\begin{table}[t!]
		\centering
		\caption{Simulation parameters}
		\label{t1}
		\begin{tabular}{ll}
			\toprule
			Parameter & Value \\
				\midrule
			Deployment area & $20$m $\times$ $20$m\\
            Subnetwork radius ($R_{sub}$) & $2$m\\
            Velocity of a subnetwork $v$ & $2$ m/s\\
            Mini-batch size $B$ & $2^M \times 30$ \\
            Time slot duration & \num {3} ms\\
            Activation probability $(\grave{p})$ & \num {0.4}\\
            Deadline threshold $(D)$ &   \num {20} time slots\\
			\bottomrule
		\end{tabular}	
	\end{table}
 We evaluate the performance of the proposed RA method via comprehensive MATLAB simulations using the parameters listed in Table~\ref{t1} (unless stated otherwise).We consider an Industrial scenario consisting of $N$ subnetworks, each comprises a set of sensors and one LAP. The subnetworks are randomly distributed in 20m $\times$ 20m deployment area, which is a typical scenario in industrial subnetworks~\cite{InF, par1}. We adopt the 3GPP TR 38.901 IIoT channel model~\cite{channel_model}, including the line-of-sight (LOS) and non-line-of-sight (NLOS) models. In addition, we employ the alpha-betagamma (ABG) model for the pathloss~\cite{3gpp38901} and spatially correlated shadowing model used in~\cite{shadow}. Essentially, the channel gain of the communication links in the system denoted as $h \in \mathcal{C}$ is a function of the path-loss $PL$, shadowing $S$ and the Rayleigh distributed small-scale fading $\zeta \sim \mathcal{CN}(0, 1)$.
\begin{equation}
    h = \zeta \times 10^{-\frac{(PL + S)}{10}}.
\end{equation}
In the simulation framework, a snapshot-based mobility model is adopted. At each snapshot, the subnetworks are randomly placed within a rectangular deployment region following a uniform spatial distribution. Each subnetwork then proceeds to move in a randomly selected direction at a constant speed $v$. The direction of motion is updated whenever a subnetwork either reaches the boundary of the area or comes within 1.5 meters of another subnetwork where this minimum separation constraint is enforced to prevent unrealistic overlaps or collisions. The learning phase of the DNN starts with more explorations than exploitations with $\epsilon$ gradually decreases from 1 to 0.1 with a step of 0.005. The learning rate in RMSProp is a hyperparameter and is gradually reduced by a decay rate of 0.015 following each alarm event. The proposed method is compared with two benchmarks. The first one is Multi-Armed Bandit-based random access(MAB-RA) which employs an $\epsilon$-greedy method to select an action~\cite{MAP-RA}.

    \begin{figure}[t!] 
		\centering
		\includegraphics[width= 0.9\linewidth]{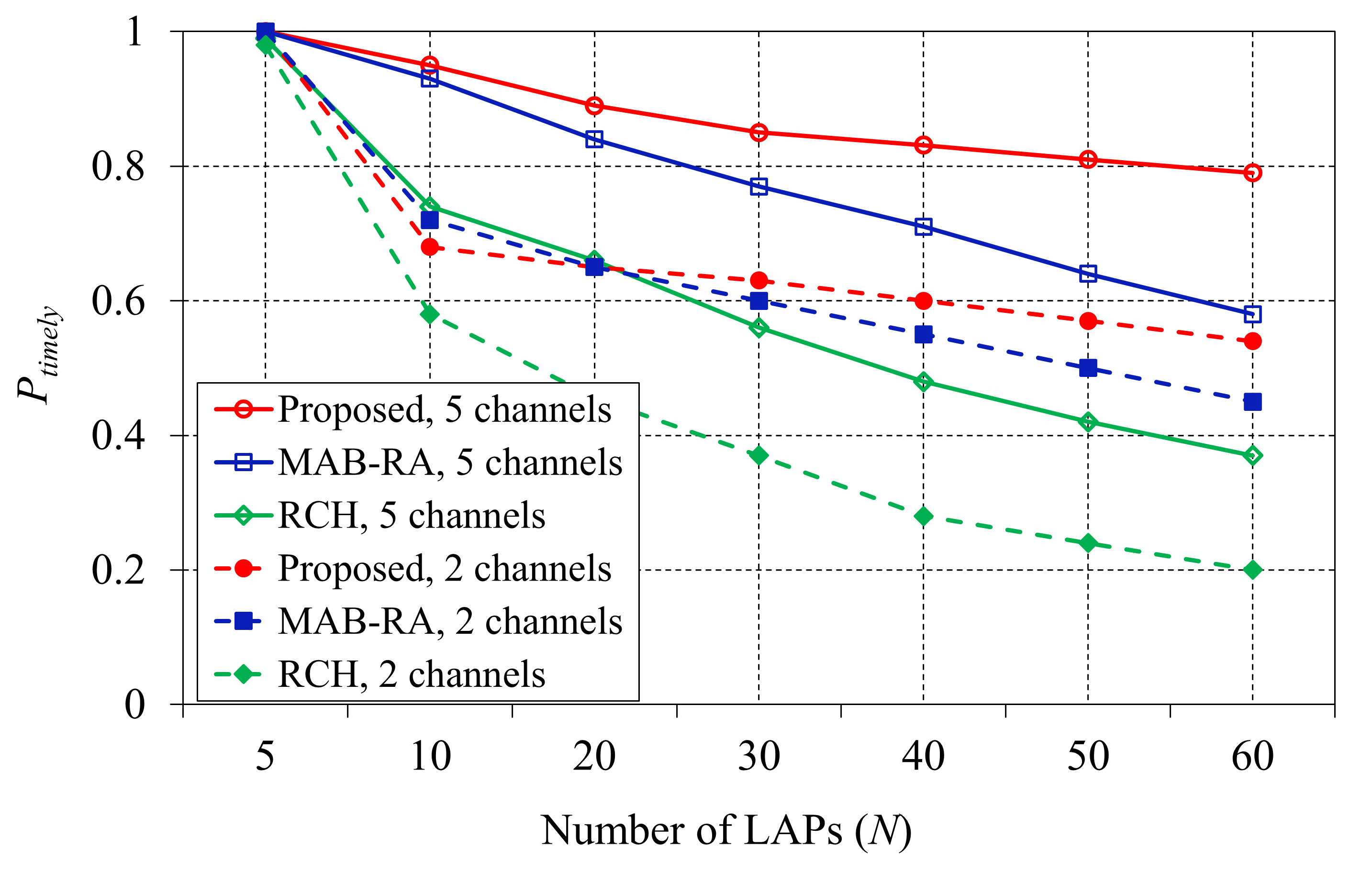}
        \vspace{-3mm}
		\caption{Performance comparison of $P_{timely}$ under different number of LAPs ($N$).  \label{LAPs}}
		\vspace{-5mm}
	\end{figure}
Figure~\ref{LAPs} illustrates the probability of timely packet delivery $P_{timely}=1-P_D$, as a function of the number of LAPs. As expected, the performance of all protocols deteriorates as $N$ increases due to intensified contention among LAPs for the shared channel resources. However, the proposed ML-based scheme consistently outperforms both the MAB-RA and  RCH  across all network sizes. While MAB-RA exhibits moderate degradation, it still relies on action-value updates that do not generalize well in high-collision regimes. In contrast, the proposed approach enables distributed LAPs to learn context-aware access configurations that reduce contention and maximize $P_{timely}$ under tight delay constraints. The performance gap becomes more pronounced as $N$ increases, highlighting the scalability and robustness of the proposed method. Specifically, with $M=5$, the proposed method has $8\%$ ($30\%$) higher $P_{timely}$ compared to MAB-RA (RCH) with $N=30$, while the gain increases to $21\%$ ($42\%$) when $N$ increases to 60.
    \begin{figure}[t!] 
		\centering
		\includegraphics[width= 0.9\linewidth]{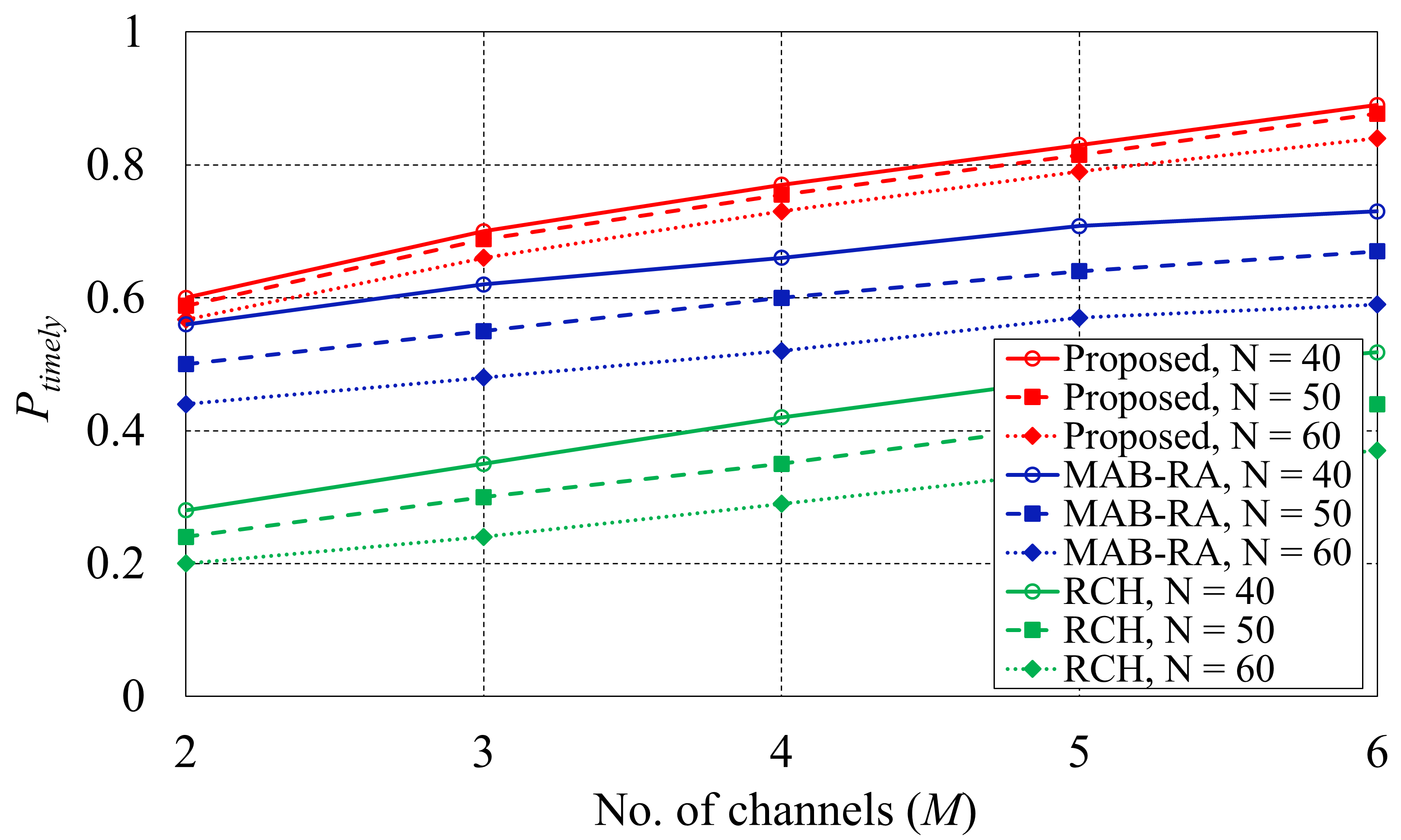}
        \vspace{-2mm}
		\caption{Performance comparison of $P_{timely}$ under different number of channels $(M)$.  \label{channels}}
		\vspace{-5mm}
	\end{figure}

In Fig.~\ref{channels}, we analyze the effect of the number of available channels $M$ on $P_{timely}$. As $M$ increases, the three methods  benefit from reduced contention, resulting in improved $P_{timely}$. Nevertheless, the proposed approach achieves a significantly higher $P_{timely}$ for all values of $M$, particularly in limited bandwidth scenarios (e.g., $M=2$), where effective coordination becomes critical. This improvement is attributed to the ability of the deep learning model to infer optimal access configurations from the CS signal, even under bandwidth constraints. With $N=60$, the proposed method gains $28\%$ improvements in $P_{timely}$ when $M$ increases from 2 to 5, while MAB-RA only experiences $13\%$ improvements. Another important note revealed from  Fig.~\ref{channels} is that the proposed method is less sensitive to the increase of the number of LAPs compared to MAB-RA and RCH.
    \begin{figure}[t!] 
		\centering
		\includegraphics[width= 0.9\linewidth]{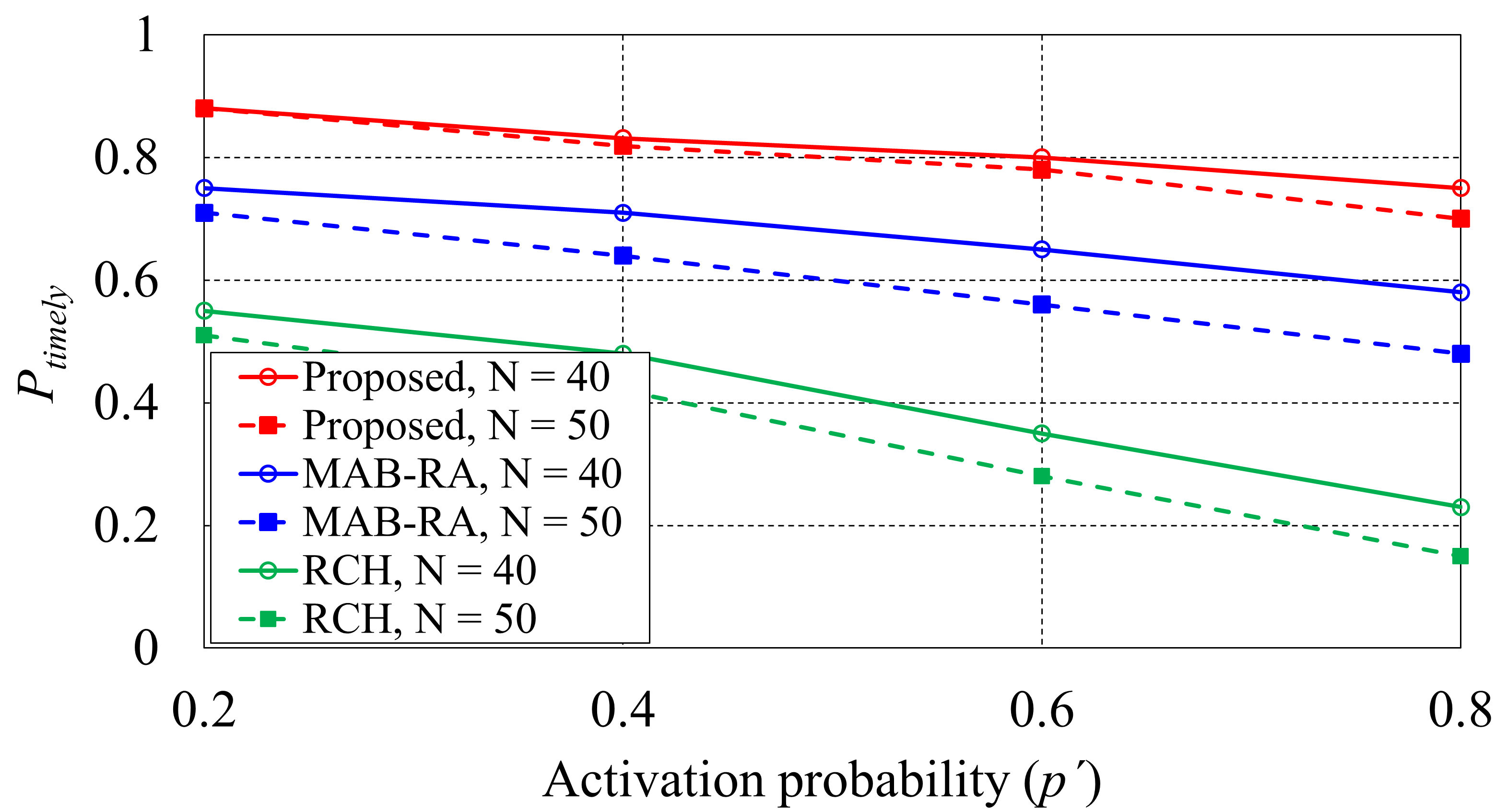}
        \vspace{-2mm}
		\caption{Performance comparison of $P_{timely}$ under different $\grave{p}$ values with $M=5$.  \label{activation}}
		\vspace{-6mm}
	\end{figure}
Figure~\ref{activation} illustrates the effect of the activation probability $\grave{p}$ on the performance of the three methods. As  $\grave{p}$ increases, the network experiences higher traffic load, resulting in increased collision likelihood and reduced $P_{timely}$ across all methods. However, we note that the proposed method maintains the lowest decay rate of $P_{timely}$ compared to MAB-RA and RCH. A notable performance gap is visible for MAB-RA and RCH when $N$ increases from 40 to 50 for the different values of $\grave{p}$, while the gap starts to be visible for the proposed method only when $\grave{p}>0.6$.


\section{Conclusion}
\vspace{-1.2mm}
\label{sec:conclusions}
This paper proposed a distributed learning-based random access protocol for time-critical communication in 6G industrial in-X subnetworks. By leveraging a broadcast contention signature signal and lightweight neural models with online training, the protocol enables implicit inter-LAP coordination without centralized control. Simulation results show significant improvements in timely packet delivery over baseline methods, particularly in dense and high-load scenarios. Future work will explore adopting the proposed coordination mechanism within goal-oriented communication frameworks for industrial control systems.
	\bibliographystyle{IEEEtran}
\bibliography{main}
	
\end{document}